\documentclass{article}
\usepackage[a4paper]{geometry}
\usepackage{amsmath}

\input{tcilatex}
\begin{document}

\title{ Generalized Geometrical Phase in the Case of Continuous Spectra}
\author{M. Maamache and Y. Saadi \\
\\
\textit{Laboratoire de Physique Quantique et Syst\`{e}mes Dynamiques,}\\
\textit{{Facult\'{e} des Sciences,Universit\'{e} Ferhat Abbas de S\'{e}tif},
S\'{e}tif 19000, Algeria}}
\date{}
\maketitle

\begin{abstract}
A quantal system in an eigenstate, of operators with a continuous
nondegenerate eigenvalue spectrum, slowly transported round a circuit $C$ by
varing parameters in its Hamiltonian, will acquire a generalized geometrical
phase factor. An explicit formula for a generalized geometrical phase is
derived in terms of the eigenstates of the Hamiltonian. As an illustration
the generalized geometrical phase is calculated for relativistic spinning
particles in slowly-changing electromagnetic fields. It is shown that the
the S-matrix and the usual scattering (with negligible reflexion) phase
shift can be interpreted as a generalized geometrical phase.

\end{abstract}

Historically, Berry's phase \cite{1} has been introduced in the context of
adiabatic evolution governed by an Hamiltonian $H\left( \vec{X}\left(
t\right) \right) $ whose parameters vary slowly in time and has been
confined to discrete spectrum.

In quantum case, the adiabatic theorem\ concerns states $\left\vert \psi
\left( t\right) \right\rangle $ satisfying the time-dependent Schr\"{o}%
dinger equation and asserts that if a quantum system with a time-dependent
non degenerate Hamiltonian $H\left( \vec{X}\left( t\right) \right) $ is
initially in the $n$th eigenstates of $H\left( \vec{X}\left( 0\right)
\right) $, and if $H\left( \vec{X}\left( t\right) \right) $ evolves slowly
enough, then the state at time $t$ will remain in the $n$th instantaneous
eigenstates of $H\left( \vec{X}\left( t\right) \right) $ up to a
multiplicative phase factor $\phi _{n}\left( t\right) $. It has been shown 
\cite{1}, that there is an additional factor $e^{\gamma _{n}^{B}}$, a part
from the familiar dynamical phase factor $e^{-\frac{i}{\hbar }\int
E_{n}\left( t\right) dt}$ associated with the time evolution of the state
being so transported with instantaneous eigenenergie $E_{n}\left( t\right) $%
, depending only the curve $C$ which has been followed in the parameters
space.

The question arises: is there a geometrical phase for a continuous spectrum?
This case was raised for the first time by R.\ G.\ Newton \cite{2} who looks
at the S matrix as geometrical phase factor. Newton \cite{2}, introduced,
what may be called the noninteraction picture to get the geometric phase
factor in the continuous spectrum. In order to reinterpret the usual
scattering phase shift as an adiabatic phase in the spirit of the original
investigation of Berry, G. Ghosh \cite{3} extends the adiabatic
approximation to the continuous spectra like an anstaz confine themselves to
one dimensional scattering with negligible reflection. Because the states
are not normalizable and gauge invariance is lost, Ghosh \cite{3}\ showed
that the phase shift can be expanded in series of reparametrization
invariant terms.

In this letter we give a generalization of the geometrical phase for the
nondegenerate continuous spectrum. Three examples are worked out for
illustration: -i) Dirac particle in a time-dependent electromagnetic field,
-ii) The explicit relation between the geometrical phase and the S matrix
with an example for application, -iii) the unidimensional reflectionless
potential

Starting with the time-dependent Schr\"{o}dinger equation governed by a
Hamiltonian $H\left( \vec{X}\left( t\right) \right) $ whose parameters vary
slowly in time%
\begin{equation}
i\hbar \frac{\partial }{\partial t}\left\vert \psi \left( t\right)
\right\rangle =H\left( \vec{X}\left( t\right) \right) \left\vert \psi \left(
t\right) \right\rangle ,  \label{SchrEq}
\end{equation}%
we define instantaneous eigenfunction in the continuous spectrum%
\begin{equation}
H\left( \vec{X}\left( t\right) \right) \left\vert \varphi \left( k,t\right)
\right\rangle =E\left( k,t\right) \left\vert \varphi \left( k,t\right)
\right\rangle .  \label{EigenVal}
\end{equation}

The adiabatic theorem demonstrated in Ref. \cite{4}, asserts that the
evolved state of an initial Hamiltonian's eigenstate at time $t_{0}$%
\begin{equation}
\left\vert \psi \left( t_{0}\right) \right\rangle =\left\vert \varphi \left(
k,t_{0}\right) \right\rangle ,  \label{IniCond}
\end{equation}%
remains at any time $t$ in the interval $\left[ k,k+\delta k\right] $%
\begin{equation}
\left\vert \psi \left( t\right) \right\rangle =\underset{k}{\overset{%
k+\delta k}{\dint }}C\left( k^{\prime };t\right) \left\vert \varphi \left(
k^{\prime };t\right) \right\rangle dk^{\prime }.  \label{EvolFunc}
\end{equation}%
Inserting (\ref{EvolFunc}) in the Schr\"{o}dinger equation (\ref{SchrEq}),
lead to%
\begin{equation}
\overset{k+\delta k}{\underset{k}{\int }}C\left( k^{\prime };t\right)
E\left( k^{\prime };t\right) \left\vert \varphi \left( k^{\prime };t\right)
\right\rangle dk^{\prime }=\overset{k+\delta k}{\underset{k}{\int }}i\hbar 
\dot{C}\left( k^{\prime };t\right) \left\vert \varphi \left( k^{\prime
};t\right) \right\rangle dk^{\prime }+\overset{k+\delta k}{\underset{k}{\int 
}}i\hbar C\left( k^{\prime };t\right) \left\vert \dot{\varphi}\left(
k^{\prime };t\right) \right\rangle dk^{\prime }  \label{Equa}
\end{equation}%
where the dot stands for time-derivative. We multiply Eq. (\ref{Equa}) by
the bra of the eigendifferential \cite{4}%
\begin{equation}
\left\vert \delta \varphi \left( k;t\right) \right\rangle =\overset{k+\delta
k}{\underset{k}{\int }}\left\vert \varphi \left( k^{\prime };t\right)
\right\rangle dk^{\prime },  \label{Bra}
\end{equation}%
this yields%
\begin{equation}
\overset{k+\delta k}{\underset{k}{\int }}C\left( k^{\prime };t\right)
E\left( k^{\prime };t\right) dk^{\prime }=\overset{k+\delta k}{\underset{k}{%
\int }}i\hbar \dot{C}\left( k^{\prime };t\right) dk^{\prime }+\overset{%
k+\delta k}{\underset{k}{\int }}C\left( k^{\prime };t\right) \left\langle
\delta \varphi \left( k;t\right) \right\vert i\hbar \frac{\partial }{%
\partial t}\left\vert \varphi \left( k^{\prime };t\right) \right\rangle
dk^{\prime }.  \label{Equa1}
\end{equation}%
Since $k$ can sweep all the possible values and the intervals $\delta k$
should be small $\left( \delta k\rightarrow 0\right) $, the equality (\ref%
{Equa1}) between integrals implies the equality between integrands, hence%
\begin{equation}
i\hbar \dot{C}\left( k^{\prime };t\right) =C\left( k^{\prime };t\right) %
\left[ E\left( k^{\prime };t\right) -\left\langle \delta \varphi \left(
k;t\right) \right\vert i\hbar \frac{\partial }{\partial t}\left\vert \varphi
\left( k^{\prime };t\right) \right\rangle \right] ,\ k^{\prime }\in \left[
k,k+\delta k\right] .  \label{Equa2}
\end{equation}%
This equation is easily integrated and gives:%
\begin{equation}
C\left( k^{\prime };t\right) =\delta \left( k^{\prime }-k\right) \exp \left[
-\underset{t_{0}}{\overset{t}{\int }}\left( \frac{i}{\hbar }E\left(
k^{\prime };t^{\prime }\right) +\left\langle \delta \varphi \left(
k;t^{\prime }\right) \right\vert \frac{\partial }{\partial t^{\prime }}%
\left\vert \varphi \left( k^{\prime };t^{\prime }\right) \right\rangle
\right) dt^{\prime }\right] ,\ k^{\prime }\in \left[ k,k+\delta k\right] ,
\label{Equa3}
\end{equation}%
hence%
\begin{equation}
\left\vert \psi \left( k,t\right) \right\rangle =\exp \left\{ \frac{i}{\hbar 
}\left[ -\gamma ^{D}\left( k;t\right) +\gamma ^{G}\left( k;t\right) \right]
\right\} \left\vert \varphi \left( k;t\right) \right\rangle ,
\label{EvolFunc1}
\end{equation}%
where $\gamma ^{D}\left( k;t\right) $ is the familiar dynamical phase
factor, and $\gamma ^{G}\left( k;t\right) $ given by%
\begin{equation}
\gamma ^{G}\left( k;t\right) =\underset{t_{0}}{\overset{t}{\int }}%
\left\langle \delta \varphi \left( k;t^{\prime }\right) \right\vert i\hbar 
\frac{\partial }{\partial t^{\prime }}\left\vert \varphi \left( k;t^{\prime
}\right) \right\rangle dt^{\prime },  \label{GeomePha}
\end{equation}%
is the generalization of the Berry's phase for the continuous spectrum. Note
that all properties of the geometrical phase in discrete case are fulfilled
by the generalized geometrical phase (\ref{GeomePha}) for the continuous
case.

As the interval $\left[ k,k+\delta k\right] $ is located inside of the
interval $\left[ -\infty ,+\infty \right] $, we can write the generalized
geometrical phase in the following practical form

\begin{equation}
\gamma ^{G}\left( k;t\right) =\underset{t_{0}}{\overset{t}{\int }}\underset{%
-\infty }{\overset{+\infty }{\int }}\left\langle \varphi \left( k^{\prime
};t^{\prime }\right) \right\vert i\hbar \frac{\partial }{\partial t^{\prime }%
}\left\vert \varphi \left( k;t^{\prime }\right) \right\rangle dt^{\prime
}dk^{\prime },  \label{GeomePhaFin}
\end{equation}%
which embodies the central result of this paper.

We now want to analyse the nature of the phase(12) through examples. Our
first case is (i) the Dirac particle in a time-dependent electromagnetic
field \cite{5, 6}, where the Dirac Hamiltonian in a 4-dimensional Hilbert
space spanned by the two-dimensional basis state $\left\vert 1\right\rangle $
and $\left\vert 2\right\rangle $, can be written 
\begin{equation}
H_{D}\left( t\right) =m\left( t\right) c^{2}\left[ \left\vert 1\right\rangle
\left\langle 1\right\vert -\left\vert 2\right\rangle \left\langle
2\right\vert \right] +c\ \sigma ^{3}\left[ p-f\left( t\right) \right]
\left\vert 1\right\rangle \left\langle 2\right\vert +c\ \sigma ^{3}\left[
p-f^{\ast }\left( t\right) \right] \left\vert 2\right\rangle \left\langle
1\right\vert ,  \label{DiracHam}
\end{equation}%
$p$ is the momentum operator, the mass $m\left( t\right) $ and the complex
parameter $f\left( t\right) $ are periodic slowly functions of time, and $%
\sigma ^{3}$ is the $2\times 2$ standard Pauli matrix. In this way $%
\left\vert \psi \left( t\right) \right\rangle $ is a 4-dimensional spinor
state. In fact, in the nonrelativistic limit \cite{5}, this Hamiltonian
reduces to a Hamiltonian of charged particle interacting with an
electromagnetic field in the so-called dipole approximation, i.e., the
vector potential $eA\left( t\right) =\func{Re}f\left( t\right) $ and the
scalar potential $eV\left( t\right) =\frac{\func{Im}f\left( t\right) }{%
2m\left( t\right) }$ are only functions of time but do not depend on
coordinates.

At any time $t$, the instantaneous \ normalized, to $\delta -$Dirac
function, eigenstates of the Hamiltonian (\ref{DiracHam}) are 
\begin{equation}
\left\vert \varphi ^{\pm }\left( z,k;t\right) \right\rangle =\left\{ \frac{%
c\sigma ^{3}g\left( k;t\right) }{\sqrt{\left[ m\left( t\right) c^{2}\mp
\hbar \omega \left( k;t\right) \right] ^{2}+\left\vert g\left( k;t\right)
\right\vert ^{2}c^{2}}}\left\vert 1\right\rangle +\frac{m\left( t\right)
c^{2}\mp \hbar \omega \left( k;t\right) }{\sqrt{\left[ m\left( t\right)
c^{2}\mp \hbar \omega \left( k;t\right) \right] ^{2}+\left\vert g\left(
k;t\right) \right\vert ^{2}c^{2}}}\left\vert 2\right\rangle \right\} \frac{%
e^{\frac{i}{\hbar }kz}}{\sqrt{2\pi \hbar }}  \label{Spinor}
\end{equation}%
corresponding, respectively, to the eigenvalues $\pm \hbar \omega \left(
k,t\right) $ $=\pm c\sqrt{m^{2}\left( t\right) c^{2}+\left\vert g\left(
k;t\right) \right\vert ^{2}}$ and $g\left( k;t\right) =f\left( t\right) -k$.

Substituting (\ref{Spinor}) in Eq. (\ref{GeomePhaFin}) leads to 
\begin{equation}
\gamma ^{G}\left( k;t\right) =\frac{i\hbar c^{2}}{2}\underset{t_{0}}{\overset%
{t}{\int }}\frac{g^{\ast }\left( k;t^{\prime }\right) \dot{g}\left(
k;t^{\prime }\right) -\dot{g}^{\ast }\left( k;t^{\prime }\right) g\left(
k;t^{\prime }\right) }{\left[ m\left( t\right) c^{2}\mp \hbar \omega \left(
k;t\right) \right] ^{2}+\left\vert g\left( k;t\right) \right\vert ^{2}c^{2}}%
dt^{\prime },  \label{Dphase}
\end{equation}%
\FRAME{ftbphFU}{6.0286in}{3.1903in}{0pt}{\Qcb{Solid angle subtended by the
circuit $C$.}}{\Qlb{SolidAngle}}{Figure}{\special{language "Scientific
Word";type "GRAPHIC";maintain-aspect-ratio TRUE;display "USEDEF";valid_file
"T";width 6.0286in;height 3.1903in;depth 0pt;original-width
5.9689in;original-height 3.1462in;cropleft "0";croptop "1";cropright
"1";cropbottom "0";tempfilename 'JZ5FSP02.wmf';tempfile-properties "XPR";}}%
when we use the following representation of the adiabatic parameters $\left( 
\func{Re}f\left( t\right) -k\right) c=\hbar \omega \sin \theta \cos \varphi $%
, $\func{Im}f\left( t\right) c=$ $\hbar \omega \sin \theta \sin \varphi $
and $m\left( t\right) c^{2}=\hbar \omega \cos \theta $; $\omega $, $\theta $
and $\varphi $ are now chosen as time-dependent adiabatic quantities, the
generalized geometrical phase is nothing but the solid angle subtended by
the circuit $C$ when seen from $\left( k,0,0\right) $ in $\left( \func{Re}%
f\left( t\right) ,\func{Im}f\left( t\right) ,m\left( t\right) c\right) $
space (see Fig.\ref{SolidAngle}), i.e.%
\begin{equation}
\gamma ^{G}\left( k;C\right) =-\frac{\hbar }{2}\Omega .  \label{SolAng}
\end{equation}

Our next example is: (ii) the geometrical aspect of the S matrix: A very
general way of looking at the S matrix as a geometric phase factor has been
implicitly provided by Newton \cite{2}. The expression for geometrical phase
given in \cite{2} looks strikingly similar to the equation of the wave
operator in the interaction picture, which leads Newton to conclude that the
S matrix appears in geometric phase as an expression of the adiabatic
switching on and off of the interaction. Here we show explicitly that, in
the case of an elastic scattering, the generalized geometrical phase (\ref%
{GeomePhaFin}) is nothing but the diagonal element of the S matrix.

The state vector $\left\vert \psi \left( t\right) \right\rangle =U\left(
t,t_{0}\right) $ $\left\vert \psi \left( t_{0}\right) \right\rangle $ of the
given physical system is assumed to satisfy the Schr\"{o}dinger equation (%
\ref{SchrEq}), $U\left( t,t_{0}\right) $ being the unitary evolution
operator associated to the Hamiltonian operator $H\left( t\right) $. In
order to solve Eq. (\ref{SchrEq}) under the adiabatic assumption, we assume
that the Hamiltonian can be split into two parts 
\begin{equation}
H\left( t\right) =H_{0}+V\left( t\right) ,  \label{Split}
\end{equation}%
so that $H_{0}$ represents the particles Hamiltonian in the absence of
interaction between them. In other words, $H_{0}$ represents the free
Hamiltonian operator. For the present we have primarily elastic scattering
in mind, and we may think of $V\left( t\right) $ as the time-dependent
potential of interaction. We assume that $V\left( t\right) $ varies slowly
in time with%
\begin{equation}
V\left( t\right) \neq 0\qquad t_{0}<t<t_{1}.  \label{Interac}
\end{equation}

In scattering problems, we are interested in calculating transition
amplitudes between states $\left\vert \varphi ^{F}\left( k\right)
\right\rangle $ (where $F$ stands for free evolution). The system initially
in the state%
\begin{equation}
\left\vert \psi \left( -\infty \right) \right\rangle =\left\vert \varphi
^{F}\left( k_{0}\right) \right\rangle  \label{InitialState}
\end{equation}%
evolves freely toward the interaction region as 
\begin{subequations}
\begin{equation}
\left\vert \psi \left( t\right) \right\rangle =\exp \left[ -\frac{i}{\hbar }%
\underset{-\infty }{\overset{t\leq t_{0}}{\int }}E\left( k_{0}\right)
dt^{\prime }\right] \left\vert \varphi ^{F}\left( k_{0}\right) \right\rangle
,  \label{BeforInterac}
\end{equation}%
under the action of the free Hamiltonian $H_{0}$ whose eigenstates in the
continuous spectrum are defined as 
\end{subequations}
\begin{equation}
H_{0}\left\vert \varphi ^{F}\left( k\right) \right\rangle =E\left( k\right)
\left\vert \varphi ^{F}\left( k\right) \right\rangle .  \label{FreeEigeVal}
\end{equation}

Let us now expand the state vector $\left\vert \psi \left( t\right)
\right\rangle $ on the basis of the instantaneous eigenstates $\left\vert
\varphi \left( k,t\right) \right\rangle $ of $H\left( t\right) $ in the
continuous spectrum 
\begin{equation}
\left\vert \psi \left( t\right) \right\rangle =\int C^{ad}\left( k;t\right)
\left\vert \varphi \left( k;t\right) \right\rangle dk,  \label{AdExpan}
\end{equation}%
(where $ad$ stands for adiabatic evolution).

Expansion of $\left\vert \psi \left( t\right) \right\rangle $ on the basis
of the instantaneous eigenstates $\left\vert \varphi ^{F}\left( k\right)
\right\rangle $ of $H_{0}$ leads to%
\begin{equation}
\left\vert \psi \left( t\right) \right\rangle =\int C^{F}\left( k;t\right)
\left\vert \varphi ^{F}\left( k\right) \right\rangle dk  \label{FrExpan}
\end{equation}%
from which we may conclude by comparison with (\ref{AdExpan}) that%
\begin{equation}
C^{F}\left( k;t\right) =\int C^{ad}\left( k^{\prime };t\right) \left\langle
\varphi ^{F}\left( k\right) |\varphi \left( k^{\prime };t\right)
\right\rangle dk^{\prime },  \label{Coef}
\end{equation}%
with the initial condition (\ref{InitialState}), i.e. $C^{F}\left( k;-\infty
\right) =\delta \left( k-k_{0}\right) $.

The coefficients $C^{F}\left( k;t\right) $ represent the matrix elements of
the evolution operator $U\left( t,t_{0}\right) $\ in the basis $\left\{
\left\vert \varphi ^{F}\left( k\right) \right\rangle \right\} $%
\begin{equation}
C^{F}\left( k;t\right) =\exp \left[ -\frac{i}{\hbar }\underset{-\infty }{%
\overset{t}{\int }}E\left( k_{0}\right) dt^{\prime }\right] \left\langle
\varphi ^{F}\left( k\right) \left\vert U\left( t,t_{0}\right) \right\vert
\varphi ^{F}\left( k_{0}\right) \right\rangle .  \label{MatrixEle}
\end{equation}

In the interaction picture, the solution of the Schr\"{o}dinger equation is
transformed to%
\begin{equation}
\left\vert \tilde{\psi}\left( t\right) \right\rangle =\exp \left[ \frac{i}{%
\hbar }H_{0}\left( t-t_{0}\right) \right] \left\vert \psi \left( t\right)
\right\rangle .  \label{InterPic}
\end{equation}

The corresponding evolution operator $\tilde{U}\left( t,t_{0}\right) $ is
related to the time-dependent unitary operator $U\left( t,t_{0}\right) $ by%
\begin{equation}
\tilde{U}\left( t,t_{0}\right) =\exp \left[ \frac{i}{\hbar }H_{0}\left(
t-t_{0}\right) \right] U\left( t,t_{0}\right) .  \label{InterOper}
\end{equation}

It is easily verified that, by using (\ref{MatrixEle}) and (\ref{InterPic}),
the matrix elements of this time-dependent operator between the eigenstates
of the unperturbed Hamiltonian $H_{0}$ satisfy 
\begin{equation}
\left\langle \varphi ^{F}\left( k\right) \left\vert \tilde{U}\left(
t,t_{0}\right) \right\vert \varphi ^{F}\left( k_{0}\right) \right\rangle
=C^{F}\left( k;t\right) \exp \left[ \frac{i}{\hbar }\left( E\left( k\right)
\left( t-t_{0}\right) +\underset{-\infty }{\overset{t_{0}}{\int }}E\left(
k_{0}\right) dt^{\prime }\right) \right] .  \label{SMatrix}
\end{equation}

Let us now insert (\ref{Equa3}) in (\ref{Coef}), this yields, taking into
account that the collision is elastic and the initial condition here is
given by (\ref{InitialState}),%
\begin{equation}
C^{F}\left( k;t\right) =\left\langle \varphi ^{F}\left( k\right) |\varphi
\left( k_{0};t\right) \right\rangle \exp \left[ -\frac{i}{\hbar }E\left(
k_{0}\right) \left( t-t_{0}\right) -\underset{t_{0}}{\overset{t}{\int }}%
\left\langle \delta \varphi \left( k_{0};t^{\prime }\right) \right\vert 
\frac{\partial }{\partial t^{\prime }}\left\vert \varphi \left(
k_{0};t^{\prime }\right) \right\rangle dt^{\prime }\right] .  \label{GeomPha}
\end{equation}

Comparison of (\ref{SMatrix}) with (\ref{GeomPha}) reveals that%
\begin{equation}
\left\langle \varphi ^{F}\left( k\right) \left\vert \tilde{U}\left(
t,t_{0}\right) \right\vert \varphi ^{F}\left( k_{0}\right) \right\rangle
=\left\langle \varphi ^{F}\left( k\right) |\varphi \left( k_{0};t\right)
\right\rangle \exp \left[ \frac{i}{\hbar }\underset{-\infty }{\overset{t_{0}}%
{\int }}E\left( k_{0}\right) dt^{\prime }-\underset{t_{0}}{\overset{t}{\int }%
}\left\langle \delta \varphi \left( k_{0};t^{\prime }\right) \right\vert 
\frac{\partial }{\partial t^{\prime }}\left\vert \varphi \left(
k_{0};t^{\prime }\right) \right\rangle dt^{\prime }\right] .
\end{equation}

As expected, the initial $\left( t\leq t_{0}\right) $ and final $\left(
t\geq t_{1}\right) $ eigenstates of the free Hamiltonian $H_{0}$ are
identical, i.e.%
\begin{equation}
\left\vert \varphi ^{F}\left( k_{0},t\geq t_{1}\right) \right\rangle
=\left\vert \varphi ^{F}\left( k\right) \right\rangle .
\end{equation}%
By pushing the initial time into the distant past i.e. letting $%
t_{0}\rightarrow -\infty $, similarly $t\rightarrow \infty $ signals that
the scattering process is complete, we thus obtain the scattering matrix or
S matrix%
\begin{equation}
\left\langle \varphi ^{F}\left( k\right) \left\vert S\right\vert \varphi
^{F}\left( k_{0}\right) \right\rangle =\left\langle \varphi ^{F}\left(
k\right) \left\vert \underset{t\rightarrow +\infty ,t_{0}\rightarrow -\infty 
}{\lim }\tilde{U}\left( t,t_{0}\right) \right\vert \varphi ^{F}\left(
k_{0}\right) \right\rangle =\delta \left( k-k_{0}\right) \exp \left[ \frac{i%
}{\hbar }\gamma ^{G}\left( k_{0};+\infty \right) \right]
\end{equation}

On the basis of this comparison we may conclude that after the Hamiltonian
completes an adiabatic circuit from $H_{0}$ via $H\left( t\right) $ back to $%
H_{0}$, the state which initially was given by $\left\vert \varphi
^{F}\left( k,t_{0}\right) \right\rangle $ has gone over into a new state
that differs from it by a unitary transformation%
\begin{equation}
S\left\vert \varphi ^{F}\left( k\right) \right\rangle =S_{k}\left\vert
\varphi ^{F}\left( k\right) \right\rangle ,  \label{SmatrixEigen}
\end{equation}%
where $S_{k}$ is the eigenvalue of the unitary S matrix and is related to
the generalized geometrical phase (\ref{GeomePhaFin}) by%
\begin{equation}
S_{k}=\exp \left[ \frac{i}{\hbar }\gamma ^{G}\left( k;+\infty \right) \right]
.  \label{SmatrixEigen1}
\end{equation}%
This result expresses the geometric aspect of the S matrix and represents
the transmission amplitude $t\left( k\right) $\ \cite{4}.

Our final example is: (iii) the unidimensional potential 
\begin{equation}
V\left( x;t\right) =-\frac{\hbar ^{2}k_{1}^{2}}{m\ \left[ \cosh \left[
k_{1}\left( x-x_{0}\left( t\right) \right) \right] \right] ^{2}}
\label{ReflessPoten}
\end{equation}%
which is reflectionless for all values of the incident energy, where $m$ is
the particle's mass, $k_{1}$ a constant and $x_{0}\left( t\right) $ is a
slowly time-dependent function satisfying\ \ $x_{0}\left( t\right) \underset{%
t\rightarrow \overset{-}{+}\text{\ }\infty }{=}\overset{+}{-}\infty .$ The
potential $V\left( x;t\right) $ (\ref{ReflessPoten}) has a single bound
state of energy $E_{k=k_{1}}=-\frac{\hbar ^{2}k_{1}^{2}}{2m}$ and whose
exact positive energy solutions given by%
\begin{equation}
\varphi \left( x,k;x_{0}\left( t\right) \right) =\left[ ik-k_{1}\tanh \left[
k_{1}\left( x-x_{0}\left( t\right) \right) \right] \right] \frac{e^{ik.x}}{%
\sqrt{2\pi }\left( k_{1}+ik\right) },  \label{RefWave}
\end{equation}%
are normalized in terms of the eigendifferentials \cite{7}.

Inserting (\ref{RefWave}) in (\ref{GeomePhaFin}) we\textbf{\ }get 
\begin{equation}
\gamma ^{G}\left( k;+\infty \right) =\frac{2\hbar k_{1}k}{k^{2}+k_{1}^{2}},
\label{GeomPhase}
\end{equation}%
which gives according to (\ref{SmatrixEigen1})%
\begin{equation}
S_{k}=t\left( k\right) =\exp \left[ 2i\frac{k_{1}k_{0}}{k_{0}^{2}+k_{1}^{2}}%
\right] =e^{i\ \delta _{0}}.  \label{TranAmplAdiab}
\end{equation}%
This agrees with the results for the transmission amplitude obtained by
Ghosh \cite{3} and is a good approximation (see fig.\ref{AmplTran}) for the
well known exact transmission amplitude \cite{3}%
\begin{equation}
t\left( k\right) =\exp \left[ 2i\arctan \left( \frac{k_{1}}{k}\right) \right]
=e^{i\delta },  \label{TranAmplExact}
\end{equation}%
obtained in the time independent case.

\FRAME{ftbphFU}{3.9911in}{2.7795in}{0pt}{\Qcb{illustrates a good agreement
between the exact value $\protect\delta $(solid line) of the argument of the
transmission amplitude and $\protect\delta _{0}$ (dashed line) the value of
the generalized geometrical phase . The interval $[-\protect\sqrt{2},\protect%
\sqrt{2}]$ corresponds to discrete spectrum region . }}{\Qlb{AmplTran}}{%
Figure}{\special{language "Scientific Word";type
"GRAPHIC";maintain-aspect-ratio TRUE;display "USEDEF";valid_file "T";width
3.9911in;height 2.7795in;depth 0pt;original-width 6.608in;original-height
4.587in;cropleft "0";croptop "1";cropright "1";cropbottom "0";tempfilename
'JZ5FSP04.wmf';tempfile-properties "XPR";}}


\begin{thebibliography}{9}
\bibitem{1} M. V. Berry, Proc. R. Soc. London, Ser. A \textbf{392}, 45
(1984).

\bibitem{2} R. G. Newton , Phys. Rev. Lett. \textbf{72}, 954 (1994).

\bibitem{3} G. Ghosh, Phys. Lett. A \textbf{210}, 40 (1996).

\bibitem{4} M. Maamache and Y. Saadi, " Adiabatic \ Theorem in the Case of
Continuous Spectra." Submited to Phys. Rev. Lett. (2008).

\bibitem{5} M. Maamache and H.Lakehal, Europhys. Lett. \textbf{67, } 695
(2004)

\bibitem{6} This example has been treated by one of us \cite{5} for the non
adiabatic case using the Lewis-Riesenfeld theory and where $f\left( t\right) 
$ is a real function of time. In this Ref. \cite{5} it was shown that, in
the adiabatic limit, a Berry's phase appears; in fact this is not true
because $f\left( t\right) $ is real which leads to a vanishing Berry's
phase. The example treated here, where $f\left( t\right) $ is a complex
parameter, gives a non vanishing geometrical phase which is the solid angle.

\bibitem{7} The functions (\ref{RefWave}) are normalized in terms of the
eigendifferentials (\ref{Bra}), indeed%
\begin{equation*}
\left\langle \varphi \left( k,x_{0}\right) |\varphi \left( k^{\prime
},x_{0}\right) \right\rangle =\delta \left( k-k^{\prime }\right) -\frac{k_{1}%
}{\pi \left( k_{1}-ik\right) \left( k_{1}+ik^{\prime }\right) }\underset{%
L\rightarrow \infty }{\lim }\cos \left[ \left( k-k^{\prime }\right) L\right]
\end{equation*}%
the last term takes any values between $-1$ and $+1$. Now, replacing $%
\left\langle \varphi \left( k,x_{0}\right) \right\vert $ by the
corresponding eigendifferential we get%
\begin{equation*}
\left\langle \delta \varphi \left( k,x_{0}\right) |\varphi \left( k^{\prime
},x_{0}\right) \right\rangle =\int_{k}^{k+\delta k}\delta \left(
k"-k^{\prime }\right) dk"+\left. \frac{k_{1}\tanh \left[ k_{1}\left(
x-x_{0}\right) \right] }{2i\pi \left( k_{1}-ik\right) \left(
k_{1}+ik^{\prime }\right) x}e^{i\left( k-k^{\prime }\right) x}\left( e^{-i\
\delta kx}-1\right) \right\vert _{-\infty }^{\infty }
\end{equation*}%
since the second term vanishes, then we can choose $\underset{L\rightarrow
\infty }{\lim }\cos \left[ \left( k-k^{\prime }\right) L\right] =0$ without
ambiguity \cite{8}.

\bibitem{8} A. Messiah, Quantum Mechanics (North-Holland 1962).
\end{thebibliography}
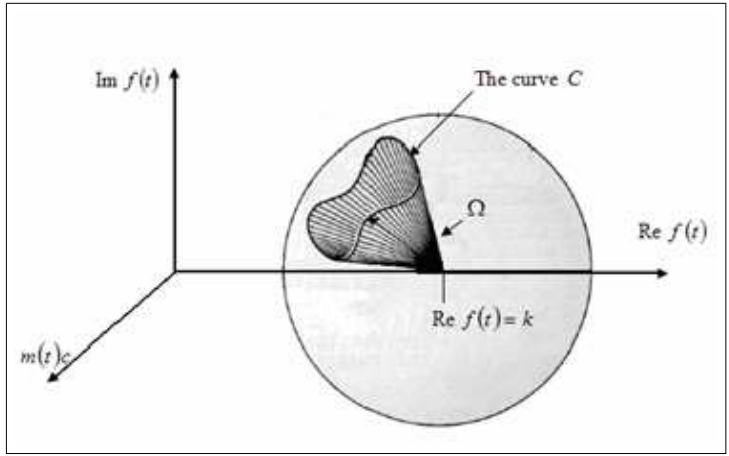
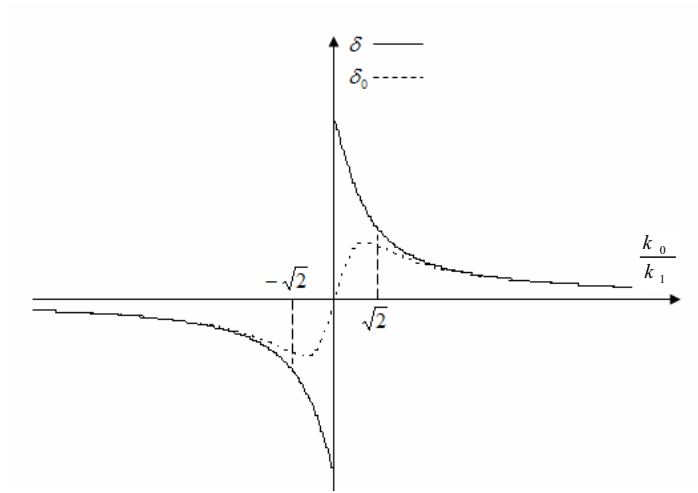
\end{document}